\documentclass[12pt,preprint]{aastex}

\def\lea{\mathrel{<\kern-1.0em\lower0.9ex\hbox{$\sim$}}}
\def\gea{\mathrel{>\kern-1.0em\lower0.9ex\h
box{$\sim$}}}

\def\etal{et al.}
\def\ha{{H$\alpha$~}}
\def\ergs{{~erg~s$^{-1}$}}
\def\h2{{\ion{H}{2}}}

\shorttitle{\h2 regions in M51}
\shortauthors{Lee, Hwang, and Lee}

\begin{document}

\title{\h2 Region Luminosity Function of the Interacting Galaxy M51}

\author{Jong Hwan Lee$^{1}$, Narae Hwang$^{2}$, and Myung Gyoon Lee$^{1}$}
\affil{$^{1}$Astronomy Program, Department of Physics and Astronomy, \\
Seoul National University, Seoul 151-747, Korea}
\affil{$^{2}$National Astronomical Observatory of Japan \\
2-21-1 Osawa Mitaka, Tokyo 181-8588, Japan}
\email{leejh@astro.snu.ac.kr, narae.hwang@nao.ac.jp, mglee@astro.snu.ac.kr}

\begin{abstract}
We present a study of \h2 regions in M51 using the Hubble Space Telescope ACS images
taken as part of the Hubble Heritage Program.
We have catalogued about 19,600 \h2 regions in M51 with \ha luminosity  
in the range of $L = 10^{35.5}$\ergs~to $10^{39.0}$\ergs.
The \ha luminosity function of \h2 regions (\h2 LF) in M51 is well represented by a double power law with 
its index $\alpha=-2.25\pm0.02$ for the bright part and $\alpha=-1.42\pm0.01$ for the faint part,
separated at a break point $L= 10^{37.1}$\ergs.
This break was not found in previous studies of M51 \h2 regions.
Comparison with simulated \h2 LFs suggests that 
this break is caused by the transition of \h2 region ionizing sources,
from low mass clusters (with $\sim 10^3$ $M_\odot$, including several OB stars) 
to more massive clusters (including several tens of OB stars).
The \h2 LFs with $L < 10^{37.1}$\ergs~are found to have different slopes for different parts in M51:  
the \h2 LF for the interarm region is steeper than those for the arm and the nuclear regions.
This observed difference in \h2 LFs can be explained by evolutionary effects
that \h2 regions in the interarm region are relatively older than those in the other parts of M51.
\end{abstract}


\section{Introduction}

\h2 regions are an excellent tracer of recent star formation
since the radiation from \h2 regions carries a signature of young OB stars.
The total flux of the hydrogen recombination line (e.g., H$\alpha$)  coming from an \h2 region 
is proportional to the total Lyman continuum emission rate of the ionizing stars,
if we assume that all ionizing photons are locally absorbed.
Thus, the \ha luminosity of the \h2 region ($L$) is an indicator of the Lyman continuum emission from ionizing sources.

In general, \h2 regions are classified into three groups,
depending on their observed \ha luminosity and size \citep{hod69, hod74, ken89, hod89, fra04}.
(a) Classical \h2 regions with \ha luminosity fainter than $L \approx 10^{37}$\ergs : 
With their sizes up to several parsecs,  these \h2 regions are typical \h2 regions ionized by several OB stars.
A representative example of this class is M42, the Orion nebula. 
(b) Giant \h2 regions with \ha luminosity of about $L = 10^{37}$\ergs~--~$10^{39}$\ergs : 
These \h2 regions are ionized by a few OB associations or massive star clusters  and usually smaller than 100 pc.
Typical examples in our Galaxy of this class are W49, NGC 3603, and the Carina nebula.
(c) Super giant \h2 regions with \ha luminosity brighter than  $L \approx  10^{39}$\ergs :
These may be ionized by multiple star clusters or super star clusters \citep{wei10}.
These \h2 regions have no known analogue in the Galaxy,
and they are mostly found in late-type galaxies or interacting systems.
Two examples of this kind are 30 Doradus in the Large Magellanic Cloud (LMC) and NGC 604 in M33.

The \ha luminosity function of \h2 regions (\h2 LF) in a galaxy
provides a very useful information on the formation and early evolution of stars and star clusters,
and has been a target of numerous studies \citep{ken80, ken89, roz96, hod99, thi02,  bra06}.
It is usually represented by a power law of the following form:

\begin{displaymath}
N(L) dL = A L^{\alpha} dL 
\end{displaymath}

\noindent where $N(L)dL$ is the number of \h2 regions with \ha luminosity between $L$ and $L+dL$,
$\alpha$ is a power law index, and $A$ is a constant. 
\citet{ken89} found that the \ha luminosity functions of bright \h2 regions in 30 nearby spiral and irregular galaxies
are represented approximately by a power law with $\alpha = -2.0\pm0.5$.
For some galaxies, \h2 LFs are known to have a break at $L \approx 10^{38.9}$\ergs,
being steeper in the bright part  than in the faint part \citep{ken89,ran92,roz96,kna98}.
The \h2 LFs with this break is called 'Type II', and this break is called Str\"omgren luminosity \citep{ken89, bra06}.
The \h2 LF slope also depends on the regions in a spiral galaxy:
the steeper \h2 LF for the interarm region than for the arm region \citep{ken89,ran92,thi00,sco01}.

Several scenarios have been suggested to explain 
the cause of the \h2 LF slope change at $L \approx 10^{38.9}$\ergs:
different molecular gas cloud mass spectra \citep{ran92},
the evolution of  \h2 regions and the variation in the number of ionizing stars \citep{von90,mck97,oey98},
the transition from ionization-bounded to density-bounded \h2 regions \citep{bec00},
and the blending of small \h2 regions caused by low resolution observations \citep{ple00}.
All these scenarios, except for one involving the observational effect,
basically suggest that the \h2 LF shape should depend on the physical conditions of
the corresponding regions in a galaxy.

Most studies on the \h2 LFs are based on ground-based images,
missing a significant fraction of faint \h2 regions \citep{ken80,ken89,bec00,thi02,bra06,hel09} 
except for a small number of the galaxies in the Local Group \citep{ken86,hod89, hod89b, hod90}.
Therefore, the nature of the faint \h2 regions in galaxies is not well known.
There are a few studies based on Hubble Space Telescope ($HST$) images,
but they covered only a small fraction of the target galaxies \citep{ple00,sco01,buc06}.
We started a project to study the \h2 regions in M51,
using the high resolution images taken with the $HST$ Advance Camera for Survey (ACS).

M51 is a system of interacting galaxies, 
including a grand-design spiral galaxy NGC 5194 (Sbc) and a barred lenticular galaxy NGC 5195 (SB0),
located in a relatively close distance (9.9 Mpc, \citealt{tik09}).
Its almost face-on orientation ($\rm i=20^{\circ}$, \citealt{tul74})
enables us to observe its structure in detail with minimal obscuration by interstellar dust.
Spiral structures in M51 have been used as a strong constraint to model the origin of spiral arms 
and the effect of galaxy interaction (\citet{dob10} and reference therein).
M51 is abundant in interstellar media \citep{sch07, hit09, egu10} 
and is well known for active star formation that has sustained at least longer than 100 Myr \citep{cal05}.
This is consistent with the existence of many young and intermediate age star clusters \citep{hwa08, sch09, hwa10, kal10}
as well as numerous \h2 regions along the spiral arms of M51 \citep{car69, van88, ken89, ran92, pet96, thi00, sco01}.

However, most of the existing studies on M51 \h2 regions are based on the ground-based observations.
In a study of 616 \h2 regions with $L > 10^{36.4}$\ergs~ in M51,
\citet{ran92} reported that the \h2 LF has an inverse slope in the  $L < 10^{37.6}$\ergs,  
and that the \h2 LF slope changes at $L \approx 10^{39}$\ergs.
He also found that the \h2 LF slope for the interarm region ($\alpha=-2.05\pm0.15$)  
is steeper than that  for the arm region ($\alpha=-1.48\pm0.07$).
\citet{pet96} estimated the \ha luminosities and sizes of 478 \h2 regions in M51
using the \ha photograph taken with the Special Astronomical Observatory 6 m telescope,  
and the \ha luminosity ranges from $L = 10^{36.4}$\ergs~to $10^{39.6}$\ergs. 
They reported a nearly flat slope below  $10^{37.6}$\ergs, in contrast to the result of \citet{ran92}, 
and  the \h2 region size distribution is well fitted by an exponential function. 

Later, \citet{thi00} found about 1,200 \h2 regions from \ha image data 
covering a central $6.7\arcmin \times 6.7\arcmin$ field of NGC 5194
using HIIPHOT that was developed for the detection and photometry of \h2 regions.
They showed that there is a change in the slope of the \h2 LF at  $L = 10^{38.8}$\ergs,
consistent with the result of \citet{ran92}.
On the other hand, \citet{sco01} detected about 1,400 \h2 regions
using the $HST$ Wide Field Camera 2 (WFPC2) observation 
covering the central $4.5\arcmin \times 4\arcmin$ field of NGC 5194.
The \h2 LF turns out to be steeper for the interarm region ($\alpha=-1.95\pm0.05$) than for the arm region ($\alpha=-1.72\pm0.03$).
However, they could not find any change in the \h2 LF slope at $L = 10^{38.8}$\ergs, in contrast to \citet{ran92} and \citet{thi00}.

These previous studies on the \h2 regions in M51 have some limitations.
The ground-based images used by \citet{ran92}, \citet{pet96}, and \citet{thi00}
cannot resolve blended \h2 regions into small or faint \h2 regions
due to its low spatial resolution of about $1.8\arcsec$ or $100$ pc at the distance of M51.
Also, it is difficult to distinguish blended low-luminosity \h2 regions from diffuse ionized gas.
On the other hand, \citet{sco01} used $HST$ image data with $0.1\arcsec-0.2\arcsec$ resolution 
($\approx 4.8-9.6$ pc at the distance of M51), which enabled to resolve blended \h2 regions and 
to distinguish faint \h2 regions from diffuse ionized gas.
However, their data cover only the central part of NGC 5194.

Recently, M51 was observed with the $HST$ ACS in \ha band as well as other continuum filters as part of the Hubble Heritage program,
and the data were released to the astronomy community \citep{mut05}.
These deep and wide field \ha imaging data provide an excellent opportunity
to study the global properties of \h2 regions in M51.
In this study, we present a result of \h2 region survey over the field covering most of
NGC 5194 and NGC 5195 regions using these data.
We make a catalog of resolved \h2 regions,
derive the \h2 LF, and investigate the properties of \h2 LF in different parts of M51.

During this study, \citet{gut10} presented a study of over 2000 \h2 regions detected 
in the same $HST$ ACS image data and reported the dependence of the mean luminosity weighted electron density 
of the \h2 regions on both \h2 region size and galactocentric radius.
\citet{gut11} presented a catalog of the 2,659 \h2 regions in M51, and showed their \ha luminosity function and size distribution. 
 
This paper is composed as follows.
\S 2 describes the data,  and \S 3 introduces the \h2 region detection and photometry procedures.
\S 4 presents  a catalog of M51 \h2 regions.
We analyze the properties of the \h2 regions, including \h2 LF and size distribution in \S 5,
and  discuss the primary results in \S 6.
Finally, a summary and conclusion is given in \S 7.

\section{Data}

M51 was observed with the $HST$ ACS in January 2005 through the Hubble Heritage program \citep{mut05}.
The data set is composed of images in four filters $F435W$ ($B$), $F555W$ ($V$), $F814W$ ($I$), and $F658N$ (H$\alpha$)
with the accumulated exposure times of 2720, 1360, 1360, and 2720 seconds, respectively.
All the necessary data reduction processes were conducted by the Hubble Heritage Team, %STScI,
including the multi-drizzling for image combination before the public release of the data.
One pixel of $HST$ ACS mosaic data corresponds to $0.05\arcsec$ after the multi-drizzling process. 
The FWHM of a point source is about $0.1\arcsec$.
The size of the field of view is $7.2\arcmin \times 10.2\arcmin$.
The spatial coverage is large enough to include the entire disk of NGC 5194 
as well as its companion NGC 5195.
Throughout this study, we adopt a distance to M51 of 9.9 Mpc 
derived using the tip of the red giant branch method  \citep{tik09}.
At this distance, one arcsecond corresponds to a linear scale of 48 pc.

\section{\h2 Region Detection and Photometry}

Detection and photometry of \h2 regions involve following steps :
(1) continuum subtraction from the \ha images,
(2) source detection and flux measurement on the continuum-subtracted images, and
(3) correction for the [\ion{N}{2}] line contamination 
and the extinction to derive net \ha fluxes of the \h2 regions.

\begin{figure}
   \plotone{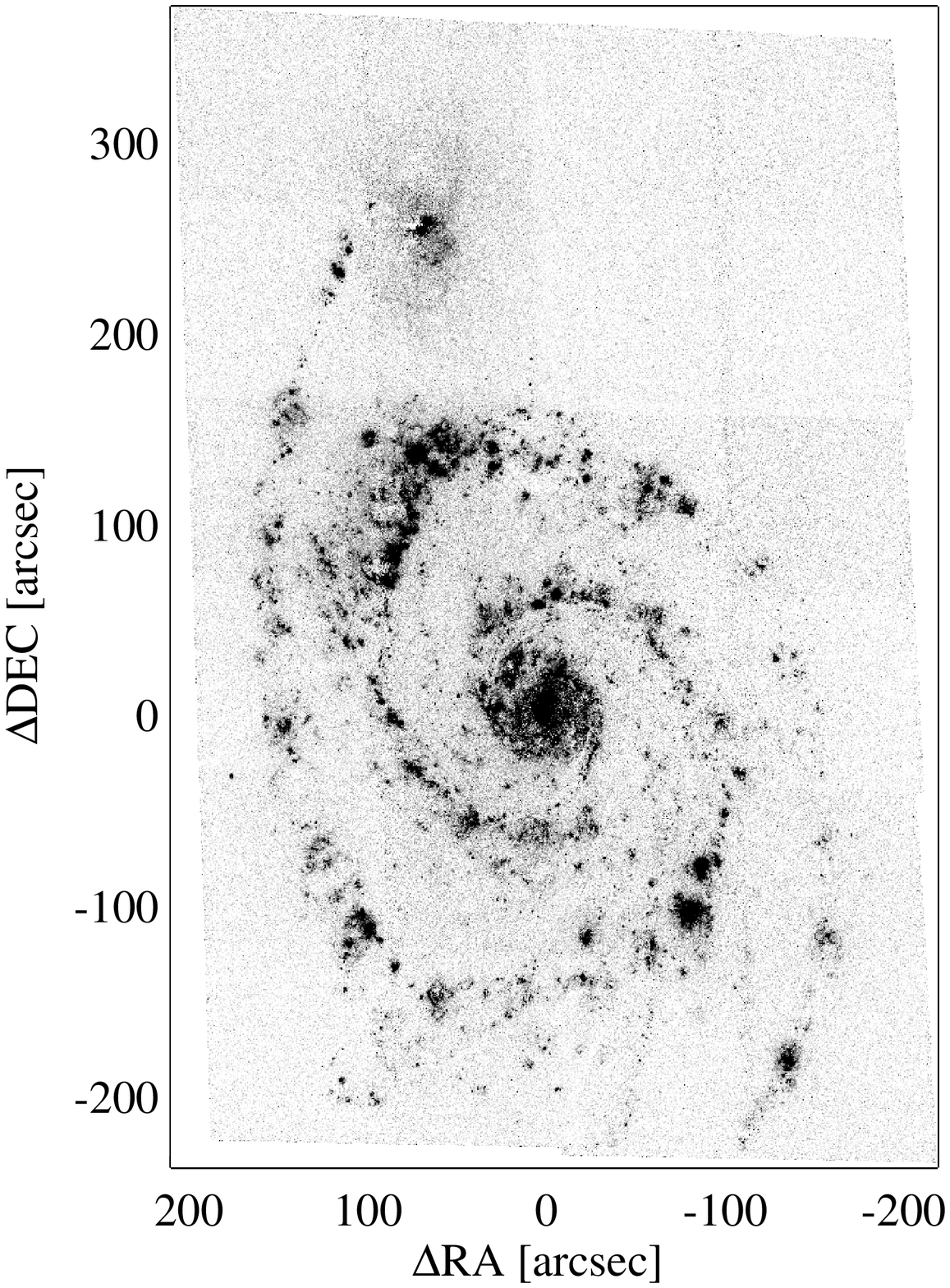}
\caption{A gray-scale map of the continuum-subtracted \ha image of M51. North is up and east to the left.
The field of view is $7.2\arcmin \times 10.2\arcmin$.
Bright discrete \h2 regions outline the spiral arms of NGC 5194,
while NGC 5195 is surrounded by faint diffuse emission.}
\label{fig1}
\end{figure}

The effective bandwidth of the $F658N$ filter is 74.8 \AA,
so that the images obtained with this filter include not only the \ha line emission but also
the [\ion{N}{2}] 6548\AA, 6583\AA~line emission and the continuum emission.
We need to subtract the continuum from the $F658N$ images
to find the \ha line emitting objects, and to measure their \ha emission line fluxes.
We used the average of $F555W$ and $F814W$ images to make a continuum image.
We measured the flux of 22 non-saturated foreground stars in the $F658N$ image
and the combined continuum image and derived a scaling factor 0.0878 from the
average ratio of these intensities.
Then, the combined continuum image was multiplied by the derived scaling factor,
and was subtracted from the $F658N$ filter image to make a 
continuum-free \ha image.
Figure \ref{fig1} displays the continuum-subtracted \ha image of M51, 
showing bright, discrete \h2 regions outlining  the spiral arms of NGC 5194, 
and the significant amount of diffuse \ha emission in the region of NGC 5195.

We detected the \h2 regions and determined their fluxes and sizes in the continuum-subtracted image 
using HIIPHOT that was developed for the detection and photometry of \h2 regions by \citet{thi00}. 
HIIPHOT determines the boundaries of individual \h2 regions
using the parameter that dictates the terminating gradient of the surface brightness profile.
We adopted a value for the terminal surface brightness gradient, 10 EM $\rm pc^{-1}$
where EM is \ha emission measure in the unit of pc $\rm cm^{-6}$.
One EM corresponds to a surface brightness of 
$2 \times 10^{-18}$ erg $\rm cm^{-2}$ $\rm s^{-1}$  $\rm arcsec^{-2}$.
The local background level for each \h2 region is obtained using a surface fit 
to pixels located within an annulus with $4\arcsec$ width outside the boundary.
We used $\rm S/N \geq 12$ as a detection limit.
With this condition we found 19,598 \h2 regions.
We calibrated the instrumental \ha fluxes ($\rm f_{H\alpha}$) of the \h2 regions 
using the photometric zero point (PHOTFLAM) given in
the header of $F658N$ image, $1.999 \times 10^{-18}$~erg~$\rm cm^{-2}$~$\rm \AA^{-1}$.
Then we derived their \ha luminosity using
$L=4\pi \rm d^2 \times f_{H\alpha}$ where $d$ is a distance to M51, 9.9 Mpc. 

We applied two corrections before obtaining the final photometry:
[\ion{N}{2}] contamination and the Galactic foreground extinction.
First, the \ha flux in the continuum-subtracted image includes contributions from the two satellite [\ion{N}{2}] lines.
At zero redshift, the [\ion{N}{2}] lines are located on either side of the 6563\AA~\ha line, one at 6548\AA~and the other at 6583\AA. 
To measure the net \ha flux, we consider the redshift $(\approx 0.002475)$ of M51, transmission values of the
$HST$ ACS $F658N$ filter, and mean [\ion{N}{2}]/\ha ratio $(\approx 0.5)$ 
obtained from spectroscopic data for ten \h2 regions in M51 by \citet{bre04}.
We used \ha / (\ha + [\ion{N}{2}]) $\approx 0.67$ to derive the net \ha flux.
Second, we applied the foreground reddening correction
to the \ha flux using the values in \citet{sch98}; E(B--V)=0.035.
For the corresponding extinction ($\rm A_{V}$ = 0.115 mag, $\rm A_{H\alpha}$= 0.092 mag),
the measured \ha flux is increased by a factor of 1.09.
Internal extinctions for individual \h2 regions are not known.
We adopted the mean extinction value $\rm A_{V} \approx 3.1$ mag derived 
from H$\alpha$/Pa$\alpha$ flux ratio for 209 \h2 regions in the central region of M51 by \citet{sco01}.
This value corresponds to a change in \ha luminosity by a factor of  $\approx 10$.

\begin{figure}
    \plotone{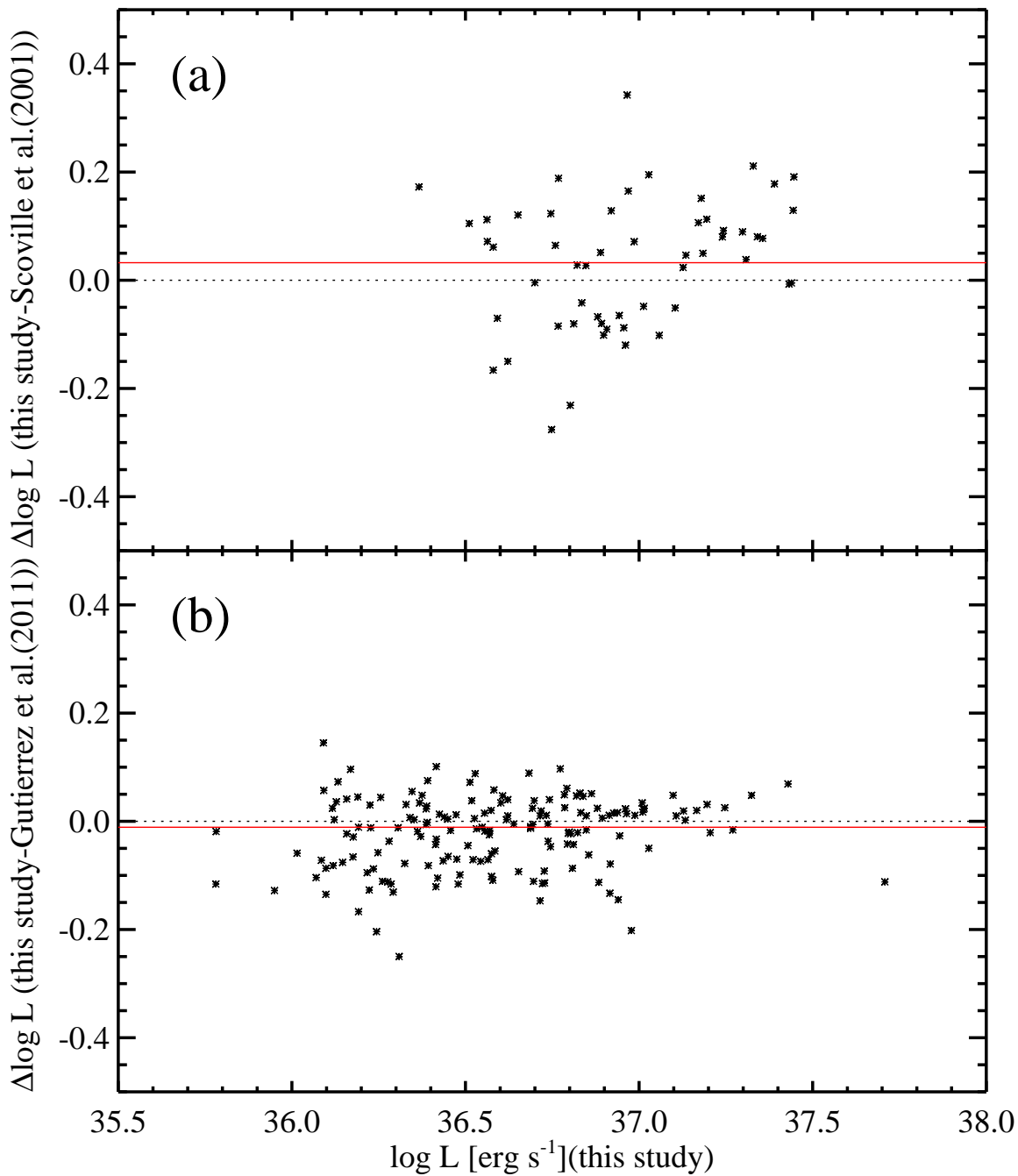}
\caption{(a) Comparison of \ha luminosity of 54 isolated compact \h2 regions common in this study and \citet{sco01}.
The mean value of the differences is $\Delta \log L$ (This study$-$\citet{sco01})= $0.03\pm0.12$ (solid line).
(b) Comparison of \ha luminosity of 164 isolated compact \h2 regions common in this study and \citet{gut11}.
The mean value of the differences is $\Delta \log L$ (This study$-$\citet{gut11})= $0.01\pm0.07$ (solid line).}
\label{fig2}
\end{figure}

%%%
There are some sets of \ha photometry data for the M51 \h2 regions available in the literature: 
\citet{ran92}, \citet{pet96}, \citet{thi00}, \citet{sco01} and \citet{gut11}.
Since the studies by \citet{ran92}, \citet{pet96}, and \citet{thi00} are based on the ground-based images,
it is difficult to compare directly our photometric measurements with theirs.
Therefore, we compared our result of \h2 region photometry with those of \citet{sco01} and \citet{gut11}.
We selected  54 isolated, compact, and bright \h2 regions common 
between this study and \citet{sco01} by visual inspection,
displaying the comparison of the measured fluxes in Figure \ref{fig2}(a).
Figure \ref{fig2}(a) shows that our photometry is in good agreement with that of \citet{sco01}.
The mean difference (log $L$ [this study] -- log $L$ [\citet{sco01}]) is $0.03 \pm 0.12$.
For Figure \ref{fig2}(b), 
we selected 164 isolated, compact, and bright \h2 regions common between this study and \citet{gut11} by visual inspection.
Figure \ref{fig2}(b) shows that our photometry is also in excellent agreement with that of \citet{gut11}.
The mean difference (log $L$ [this study] -- log $L$ [\citet{gut11}]) is $0.01 \pm 0.07$.

\section{\h2 Region Catalog}

Measured properties of \h2 regions in a galaxy are dependent on
both the resolution of images and the procedures used to define the boundaries of \h2 regions.
For example, higher resolution images can resolve some  large \h2 regions into multiple components,
reducing the number of the detected large \h2 regions.
In addition, the criteria adopted to separate the \h2 regions from the background are critical.
In this study, we present two \h2 region catalogs: for the original and group samples.
The original sample includes \h2 regions detected and provided from HIIPHOT, 
and the group sample includes \h2 regions produced by combining smaller ones in close proximity into one large one.

The original sample catalog includes about 19,600 \h2 regions,
containing  resolved sources in blended \h2 regions as well as  isolated \h2 regions.
This catalog is useful for the  study of the relation between classical \h2 regions and star clusters
since most \h2 regions in the original sample are classified as classical \h2 regions,
and only ten percent of them are giant \h2 regions.
There is no super giant \h2 regions in the original sample.
In Table \ref{table1}, we list positions, \ha luminosities, 
and effective diameters (derived from $D=2\times(\rm area/\pi)^{1/2}$).
Table \ref{table1} lists a sample of \h2 regions for reader's guide 
and the full catalog will be available electronically from the Online Journal.

To prepare the group sample, we found associations of \h2 regions in the original sample 
using the Friends-of-Friends algorithm (FOF, \citealt{dav85}).
Every source in groups has a friend source within a distance smaller than some specified linking length. 
If we increase the linking length, the number of members in individual groups increases,
decreasing the number of groups. 
As a test, we made 20 group sample sets using the linking length 
ranging from 0.1 to 2.0 mean separation length with a step of 0.1.
Through this test, we found that the linking length of 1.0 mean separation 
is the most suitable in reproducing large \h2 regions found in ground-based images.

We call these associations of \h2 regions as grouped \h2 regions.
The total number of the grouped \h2 regions is 2,294.
We inspected these grouped \h2 regions on the continuum-subtracted \ha image,
revised the membership of 209 \h2 regions.
The final number of grouped \h2 regions is 2,296,
and the number of the isolated \h2 regions which do not belong to any group is 4,919.
Thus, the sum of these two kinds of \h2 regions is 7,215.
The position of a grouped \h2 region is derived from
average values of the positions of all individual \h2 regions in a group.
The area and luminosity of a grouped \h2 region are derived from 
summing the values of all individual \h2 regions in a group.
Most of the 2,296 grouped \h2 regions are giant \h2 regions with $L > 10^{37}$\ergs,
and 18 of them are super giant \h2 region with $L > 10^{39}$\ergs. 
Table \ref{table2} lists positions, \ha luminosities, and effective diameters of 
\h2 regions in the group sample.
The last column in Table \ref{table1} marks the group ID numbers.

We compared positions of 7,216 \h2 regions in the group sample 
with those of 2,659 \h2 regions in \citet{gut11}.
2,402 HII regions in the Gutierrez et al's catalog were found in our catalog, showing that about 90 percent of the  Gutierrez et al's catalog is included in our catalog.
Among the \h2 regions in \citet{gut11} not matched with our \h2 regions,
63 \h2 regions were included as part of the grouped \h2 regions in the group sample,
and 95 \h2 regions were splitted into multiple \h2 regions.
The genuine membership of these \h2 regions is very difficult to determine.
20 \h2 regions are located outside the field used in our study.
There are 79 \h2 regions not found in our survey.
We inspected carefully the continuum-subtracted \ha image of these \h2 regions,
and found (a) that there is no recognizable \ha emission for 53 \h2 regions,
and (b) that 26 \h2 regions are located in the faint diffuse emission feature,
but with lower S/N than used in our survey.

\section{Results}

\subsection{\ha Luminosity Function of \h2 Regions}

The total number of M51 \h2 regions listed in the original sample is about 19,600.
The \ha luminosity of these \h2 regions ranges from $L = 10^{35.5}$\ergs~to $10^{39.0}$\ergs.
This detection limit is similar to the deep survey of \h2 regions in the dwarf galaxies of the Local Group 
\citep{ken80, hod89, hod89b, hod90}.
The total \ha luminosity of all these \h2 regions is derived to be  $L = 10^{41.2}$\ergs.  
There are 3,653 \h2 regions with $L > 10^{37}$\ergs, 
and their total luminosity corresponds to 70 percent of the total \ha luminosity of M51. 
There are only 160 \h2 regions with $L > 10^{38}$\ergs.

We derived the \h2 LF for the original sample, as shown in Figure \ref{fig3}.
The lower limit of \ha luminosities is determined by the observational threshold on the surface brightness.
The \h2 LF shows a maximum at $L \approx 10^{36.1}$\ergs~and declines rapidly as the luminosity decreases. 
This value is similar to the one for M33 \h2 regions, $L \approx 10^{35.9}$\ergs,
found from the completeness-corrected \h2 LF by \citet{wyd97}, 
and those for NGC 6822 and the Magellanic Clouds, $L \approx 10^{36}$\ergs \citep{hod89, wil91}.
Correcting this value for the adopted mean extinction value for M51 \h2 regions leads to $\approx 10^{37.1}$\ergs. 
This value is similar to that for M42, $\approx 10^{36.9}$\ergs \citep{sco01}. 
Considering that the \h2 LF for M33 shows a flattening for $L = 10^{35}$\ergs $-$ $10^{36}$\ergs \citep{hod99}, 
the rapid decline for $L < 10^{36}$\ergs~in our \h2 LF appears to be due to incompleteness in our catalog. 

The \h2 LF appears to be fitted by a double power law with a break at $L \approx 10^{37.1}$\ergs.
We used a double power law function to fit the \h2 LF:
\begin{displaymath}
N(L)dL=AL^{\alpha_{1}}dL\qquad \alpha_{1}: {\rm Bright~power~law~index \qquad for}~L\geq{L}_{b},
\end{displaymath}  and
\begin{displaymath}
N(L)dL=A' L^{\alpha_{2}}dL\qquad \alpha_{2}: {\rm Faint~power~law~index \qquad for}~L<{L}_{b}
\end{displaymath}
\noindent where $L_{b}$ is the break point luminosity, and  $A'=A L_{b}^{(\alpha_{2}-\alpha_{1})}$.
For the bright \h2 regions ($10^{37.1}$ \ergs~$< L <10^{38.8}$ \ergs), we obtained a power law index, $\alpha=-2.25\pm0.02$. 
For the faint part ($10^{36.1}$\ergs~$< L < 10^{37.1}$\ergs), 
the \h2 LF becomes flatter than for the bright part, having $\alpha=-1.42\pm0.01$.  

A distinguishable point with this \h2 LF is a break at $L=10^{37.1}$\ergs.
This break has not been observed in previous studies of the \h2 regions in M51, 
because the data used in previous studies were not deep enough to detect it \citep{ran92, pet96, thi00, sco01}.
However, the \h2 LF break at similar luminosity is known to exist for some nearby galaxies:
M31 \citep{wal92} and M33 \citep{hod99}.
This break may be caused by the transition of \h2 region ionizing sources, 
from low mass clusters (including several OB stars) 
to more massive clusters (including several tens of OB stars) \citep{ken89,mck97,oey98}. 
This will be discussed in detail in the next session.

We compared the \h2 LF for the original sample (solid line) with that in \citet{thi00} (dashed line),
as shown in Figure \ref{fig4}(a).
The \ha luminosity of \h2 regions for the original sample ranges from $L = 10^{35.5}$\ergs~to $10^{39.0}$\ergs,
while it ranges $L = 10^{36.4}$\ergs~to $10^{39.4}$\ergs~in \citet{thi00}.
The minimum luminosity of the \h2 regions in the original sample is about ten times fainter than that in \citet{thi00}.
While this study shows a break at $L = 10^{37.1}$\ergs,
the study by \citet{thi00} shows a break at $L=10^{38.8}$\ergs, much brighter than that in this study.   
The power law indices in this study are $\alpha=-2.28\pm0.02$ for the bright part 
 and $\alpha=-1.38\pm0.01$ for the faint part,
and those in \citet{thi00} are $\alpha=-3.41\pm0.08$ and $\alpha=-1.61\pm0.03$. 

Figure \ref{fig4}(b) shows the \h2 LF for the original sample (solid line) and that in \citet{sco01} (dashed line).
The spatial coverage of \ha images used in this study
is large enough to include most of the entire disk of NGC 5194 as well as its companion NGC 5195.
However, \citet{sco01} obtained the \h2 LF from \ha images covering only the central $4.5\arcmin \times 4\arcmin$ field of NGC 5194.
For comparison, we plotted two \h2 LFs derived for the same area.
The minimum and maximum luminosities of the \h2 regions in two samples are similar,
but there is a difference in the numbers of \h2 regions with $L < 10^{37.1}$\ergs.
The ACS images used in this study have longer exposure time than that of the WFPC2 images used in \citet{sco01},
so that we found many more faint \h2 regions than \citet{sco01}.
Both \h2 LFs are fitted by a power law for the bright part ($10^{37.0}$\ergs~$< L < 10^{38.6}$\ergs).
The \h2 LF for the original sample is slightly steeper with $\alpha=-2.24\pm0.03$ than
the \h2 LF with $\alpha=-1.91\pm0.04$ reported by  \citet{sco01} .

%%%
Figure \ref{fig4}(c) shows the \h2 LF for the original sample (solid line) and that in \citet{gut11} (dashed line).
The \ha luminosity of the \h2 regions for the original sample ranges from $L = 10^{35.5}$\ergs~to $10^{39.0}$\ergs,
while it ranges $L = 10^{35.8}$\ergs~to $10^{40}$\ergs~in \citet{gut11}.
While this study shows a break at $L = 10^{37.1}$\ergs, 
the study by \citet{gut11} shows a break at $L=10^{38.7}$\ergs, much brighter than that in this study.   
The power law indices in this study are $\alpha=-2.21\pm0.02$ for the bright part 
and $\alpha=-1.41\pm0.02$ for the faint part,
and those in \citet{gut11} are $\alpha=-2.29\pm0.17$ and $\alpha=-1.67\pm0.03$. 
This large discrepancy between the two studies is primarily due to two factors: 
(a) we considered the small local peaks for the large \h2 regions in \citet{gut11} as separated \h2 regions in this study. 
(b) we found many isolated faint \h2 regions, missed in \citet{gut11}.

In Figure \ref{fig5}(a) we compared the \h2 LF for the group sample (solid line) with that for the original sample (dashed line).
Although both \h2 LFs are fitted by a double power law, 
the location of a break and the power law indices are different.
The locations of the break are $L = 10^{37.1}$\ergs~and $10^{38.8}$\ergs, respectively.
The power law indices for the bright part are $\alpha=-2.21\pm0.02$ and $-2.60\pm0.21$,
and those for the faint part are  $\alpha=-1.41\pm0.02$ and $-1.69\pm0.01$
for the original and group samples, respectively. 

%%%
Figure \ref{fig5}(b) shows the \h2 LFs for the grouped \h2 regions (dotted line), the group sample (solid line) 
and the \h2 regions in \citet{thi00} (dashed line).
The locations of the break and the power law indices for the latter two samples are similar. 
The locations of the break for both the group sample and \citet{thi00} are $L = 10^{38.8}$\ergs.
The power law indices for the group sample are $\alpha=-2.60\pm0.21$ for the bright part 
and $\alpha=-1.69\pm0.01$ for the faint part,
and those in \citet{thi00} are $\alpha=-3.41\pm0.08$ and $\alpha=-1.61\pm0.03$. 
In conclusion, the \h2 LF break seen at $L = 10^{38.8}$\ergs~ 
in the previous studies based on the ground-based images \citep{ran92,pet96,thi00} 
can be explained by the blending of multiple \h2 regions.

In Figure \ref{fig5}(c) we compared the \h2 LFs for the group sample (solid line), the grouped \h2 regions (dotted line), 
and the \h2 regions in \citet{gut11} (dashed line). 
The \h2 LFs for the latter two samples show nearly the same shape. 
The locations of the break and the power law indices for the group sample and in \citet{gut11} are similar. 
The location of the break for the group sample and that in \citet{gut11} are $L = 10^{38.8}$\ergs~and $10^{38.7}$\ergs, respectively.
The power law indices for the group sample are $\alpha=-2.60\pm0.21$ for the bright part 
and $\alpha=-1.69\pm0.01$ for the faint part,
and those in \citet{gut11} are $\alpha=-2.29\pm0.17$ and $\alpha=-1.67\pm0.03$. 
Both studies show a good agreement at $L > 10^{37.4}$\ergs,
however, there is a large difference in the number distribution of \h2 regions at $L < 10^{37.4}$\ergs.  
It is primarily due to the existence of many isolated faint \h2 regions that were found in our survey,
but missed in \citet{gut11}.

\subsection{Spatial Variation of \ha Luminosity Function for \h2 Regions}

We investigated the spatial variation of the \h2 LF for the original sample. 
We separated \h2 regions in the original sample into four groups, according to their positions in the galaxy:
arm, interarm, and nuclear regions of NGC 5194, and NGC 5195.
The positions of the \h2 regions are listed in Table \ref{table1}.
There are 12,245 \h2 regions in the arm region, 4,422 \h2 regions in the nuclear region,
and 2,639 \h2 regions in the interarm region of NGC 5194.
We found only 292 \h2 regions in the region of NGC 5195.
The total luminosity of the arm \h2 regions is derived to be $L = 10^{41.1}$\ergs~
and that of the nuclear \h2 regions is $L = 10^{40.6}$\ergs,
while that of the interarm \h2 regions is only $L = 10^{40.1}$\ergs,
8 percent of the total \ha luminosity of all NGC 5194 \h2 regions.

Figure \ref{fig6}(a) shows the \h2 LFs for the arm (solid line), interarm (dashed line), and nuclear (dotted line) regions of NGC 5194.
The minimum luminosities ($L\approx10^{35.5}$) in the \h2 LFs are nearly the same in three regions,
but the maximum luminosities are different depending on the position.  
The maximum luminosities for the arm and nuclear regions are $L = 10^{39.0}$\ergs~and $L = 10^{38.8}$\ergs, respectively.
However, there is no \h2 region brighter than $L = 10^{38.2}$\ergs~for the interarm region.

These \h2 LFs are fitted by a double power law with a break at $L = 10^{37}$\ergs.
The power law indices for the bright part are $\alpha=-1.92\pm0.03,-2.01\pm0.09$, and $-2.08\pm0.05$
for the arm, interarm, and nuclear regions, respectively.
On the other hand, the power law indices for the faint part are $\alpha=-1.33\pm0.01,-1.53\pm0.05$, and $-1.33\pm0.03$. 
Although the \h2 LFs for the bright part ($L > 10^{37}$\ergs) have similar power law indices in three regions,
the \h2 LF for the faint part ($L < 10^{37}$\ergs) is steeper in the interarm region than in the arm and nuclear regions.

%%%
Figure \ref{fig6}(b) and (c) shows the spatial variation of the \h2 LFs 
for the grouped \h2 regions and the group sample, respectively. 
The \h2 LFs for the grouped \h2 regions in Figure \ref{fig6}(b) are truncated at $L = 10^{37}$\ergs.
We used a single power law to fit the \h2 LFs for the range from  $L = 10^{37.0}$\ergs~to $L = 10^{38.4}$\ergs. 
The power law indices are $\alpha=-1.57\pm0.04,-1.74\pm0.08$, and $-1.45\pm0.06$ 
for the arm, interarm, and nuclear regions, respectively.
The \h2 LF is steeper in the interarm region than in the arm and nuclear regions.
On the other hand, the \h2 LFs for the group sample in Figure \ref{fig6}(c) are truncated near $L = 10^{36}$\ergs.
The power law indices are $\alpha=-1.78\pm0.04,-1.95\pm0.08$, and $-1.58\pm0.06$ 
for the arm, interarm, and nuclear regions, respectively.
The \h2 LF is also steeper in the interarm region than in the arm and nuclear regions.

\subsection{Size Distribution of \h2 Regions}

The size distribution of the \h2 regions in the original sample is displayed in Figure \ref{fig7},
showing that their sizes (diameters) range from 8 pc to 110 pc.
Most of the previous \h2 region studies show that the cumulative size distribution of the \h2 regions in a galaxy 
can be well fitted with an exponential function \citep{van81,hod87,cep90,kna98,hak07}.
However, some other studies suggested that the power law form fits the differential size distribution
better than the exponential function \citep{ken80,elm99,ple00,oey03,buc06}.

Figure \ref{fig7}(a) and (b) display the cumulative and differential size distributions of the \h2 regions, respectively.
The cumulative size distribution of the \h2 regions with 15 pc $< D <$ 80 pc is fitted well 
with an exponential law:  $N(D)=N_{0}$ exp $(-D/D_{0})$ where $N_{0}=77,880$ and $D_{0}=9.3$ pc. 
The differential size distribution of these \h2 regions is fitted by a double power law with a break at $D=30$ pc.
The power law index for small \h2 regions with 15 pc $< D <$ 30 pc is $\alpha_D =-1.78\pm0.04$,
whereas $\alpha_D=-5.04\pm0.08$ for large \h2 region with 30 pc $< D <$ 110 pc.
The small \h2 regions with $D < 30$ pc may be ionized by several OB stars. 
On the other hand, the large \h2 regions with $D > 30$ pc are considered to be superpositions of small \h2 regions,
and they can be associated with stellar associations including several tens of OB stars.

Figure \ref{fig8}(a) shows the \h2 LF for small \h2 regions with $D < 30$ pc.     
Most of the small \h2 regions with $D < 30$ pc are fainter than $L = 10^{37.1}$\ergs,
and only 1,065 \h2 regions are brighter than $L = 10^{37.1}$\ergs~(solid line).
Out of the \h2 regions with $D < 30$ pc, we selected 4,320 \h2 regions 
which are neither blended with neighbor sources nor located in the crowded regions (dashed line).
Most of them are in the range from $L = 10^{35.4}$\ergs~to $10^{37.2}$\ergs.
The maximum $L$ for the isolated \h2 regions corresponds to about ten times of M42,
with an extinction-corrected value, $L \approx 10^{36.9}$\ergs.
On the other hand, Figure \ref{fig8}(b) shows the \h2 LF for large \h2 regions with $D > 30$ pc.
There are 1,808 \h2 regions brighter than $L = 10^{37.1}$\ergs,
and 632 \h2 regions fainter than $L = 10^{37.1}$\ergs.

%%%
We compared the size distribution of the \h2 regions in this study with those in \citet{thi00}, \citet{sco01} and \citet{gut11}.
Figure \ref{fig9}(a) plots the size distribution of the \h2 regions in the original sample (solid line),
the group sample (dashed line) and \citet{sco01} (dot-dashed line).
In comparison with \citet{sco01}, we found that there are many more small \h2 regions in the original sample.
We used a single power law to fit the size distributions of \h2 regions for the range from 30 pc to 125 pc. 
The power law indices are  $\alpha_D =-5.94\pm0.26$, $-3.23\pm0.10$, and $-2.79\pm0.06$
for the original sample, the group sample, and \citet{sco01}, respectively. 
Thus the size distribution of the original sample is much steeper than that of the group sample 
and that of \citet{sco01} sample, while the latter two samples have similar power law indices.

Figure \ref{fig9}(b) shows the size distribution of the grouped \h2 regions (solid line),
the \h2 regions in the group sample (dashed line), in \citet{thi00} (dotted line) and in \citet{gut11} (dot-dashed line).
The sizes of the \h2 regions in the group sample range from 8 pc to 310 pc,
while those of the \h2 regions in \citet{thi00} range from 80 pc to 480 pc.
This large discrepancy between the two studies is considered to come from the different spatial resolutions.
We found that there are many more small \h2 regions in the group sample, comparing with \citet{gut11}.
In comparison between the grouped \h2 regions and the \h2 regions in \citet{gut11},
we found that although the size distribution for the small \h2 regions ($D < 50$ pc) have similar shapes, 
the size distribution for the large \h2 regions ($D > 50$ pc) shows different shapes, having a steeper power slope 
for the grouped \h2 regions. 
The power law indices are $\alpha_D =-3.48\pm0.08$ and $-2.75\pm0.11$ 
for the grouped \h2 regions and in \citet{gut11}, respectively. 
In other words, the size of the \h2 regions derived from this study is generally smaller than in \citet{gut11}.
It is because we derived the sizes of the grouped \h2 regions 
from summing the numbers of pixels contained in all individual \h2 regions in a group
but \citet{gut11} considered the substantial empty volumes to measure the size of \h2 regions. 

\section{Discussion}

\subsection{Model Predictions for \ha Luminosity Functions}

In order to understand the nature of the observational \h2 LF,
we performed Monte Carlo simulations of \h2 LFs, following \citet{oey98}.
The \ha luminosity of an \h2 region depends on (a) the stellar initial mass function (IMF) and
(b) the relation of Lyman continuum emission rate $(\rm Q_{Lyc})$ and stellar mass.
We adopted a Salpeter IMF \citep{sal55} with $N(m_{*})dm_{*}\propto {m_{*}}^{-2.35}$
over the range from 1 $\rm M_{\odot}$  to 100 $\rm M_{\odot}$.
We used $\rm Q_{Lyc}$ depending on stellar mass given by \citet{vac96}
which is based on stellar models with constraints provided by observations of individual high-mass stars.

We considered artificial clusters with the number of stars per cluster ($\rm N_{*}$) ranging from 1 to 10,000,
and estimated the total $\rm Q_{Lyc}$ radiating from each cluster.
To reduce the statistical noise, we repeated the experiment  10,000 times, 
and derived the average $\rm Q_{Lyc}$ of  the individual cluster with a specific $\rm N_{*}$. 
Then we obtained the \ha luminosity of this cluster using $L = 1.37 \times 10^{-12}~\rm Q_{Lyc}$ given in \citet{sco01}. 

We represented the distribution of numbers of stars in each model cluster by a power law:
$N(N_{*})~dN_{*}$ = ${N_{*}}^{\beta}dN_{*}$  
where $N(N_{*})dN_{*}$ is the number of stars per cluster in the range from $N_{*}$ to $N_{*}+dN_{*}$,
and $\beta$ is a power law index.
We assumed that the star formation in each cluster occurs on a short time scale compared
with stellar evolution time scales, and that all clusters in a galaxy are created at the same time.
We converted simulated \h2 LFs into observational \h2 LFs for comparison,
considering the internal extinction as adopted above.

The shape of the \h2 LFs is determined by two parameters: $\beta$ 
and the maximum number of stars per cluster, $\rm N_{*,max}$.
To know how these two parameters affect the \h2 LF,
we constructed nine \h2 LFs with 300,000 zero-age clusters  
for three power law indices ($\beta = -1.60$, --2.00, and --2.40)
and three values of $\rm N_{*,max}$ (330, 1,000, and  10,000), plotting them in Figure \ref{fig10}.
The mass of a cluster with $\rm N_{*,max}=330$ and $\beta=2.0$ is 1,000 $M_\odot$ and 2,500 $M_\odot$, respectively,
for the stellar mass range of 1--100 $M_\odot$ and 0.1--100 $M_\odot$.
The mass of a cluster with $\rm N_{*,max}=1,000$ is 3,090 $M_\odot$ and 7,700 $M_\odot$, 
and the mass of a cluster with $\rm N_{*,max}=10,000$ is 30,900 $M_\odot$ and 77,000 $M_\odot$,
for the same ranges of stellar mass.
The number of OB stars with $M>17 M_\odot$ 
$\sim 7$, $\sim 20$ and $\sim 60$ for $\rm N_{*,max}=330, 1,000$ and 10,000, respectively.

The \h2 LFs with $\rm N_{*,max} = 330$ in the upper panels of Figure \ref{fig10} are truncated at $L = 10^{37.0}$\ergs.
Clusters with $\rm N_{*} < 330$ produce few \h2 regions brighter than $L = 10^{37.0}$\ergs.
They show nearly flat slopes  for $\beta = -1.60$ and --2.00, but show some slope for $\beta=-2.40$.
In the middle panels, the \h2 LFs with $\rm N_{*,max} = 1,000$ show the existence of some bright \h2 regions with $L > 10^{37.0}$\ergs.
They show a change in slope at  $L \approx  10^{37.0}$\ergs,
which is similar to the value found for M51 \h2 regions in this study (Figure \ref{fig3}).
We fit the bright part ($L > 10^{37.0}$\ergs) with a power law, 
obtaining $\alpha=-1.48$, --1.86, and --2.23 for $\beta=-1.60$, --2.00, and --2.40, respectively.
Thus the values of $\alpha$ decreases as $\beta$ decreases.
The \h2 LFs with $\rm N_{*,max} = 10,000$ in the low panels show a more obvious break at $L =  10^{37.0}$\ergs. 
Fitting the bright part ($L >  10^{37.0}$\ergs) yields 
$\alpha=-1.55$, --1.94, and --2.36 for $\beta=-1.60$, --2.00, and --2.40, respectively.
Thus the values of $\alpha$ are very similar to those for $\beta$, 
showing that the values of these two parameters get similar as $\rm N_*$ increases.

For comparison of the \h2 LFs in Figure \ref{fig10} with the observational ones,
we need to consider the effect of evolution of $\rm Q_{Lyc}$.
We assumed that $\rm Q_{Lyc}$ remains constant during the main-sequence lifetimes and is zero thereafter.
Main-sequence lifetimes ($\rm t_{ms}$) for individual OB stars are obtained from \citet{mae87}.
Figure \ref{fig11} shows the \h2 LFs with $\beta =$ --2.40  and $\rm N_{*,max} =$ 10,000 
derived for four ages 0, 2, 4, and 6 Myr.
We adopted a value for $\beta$, similar to the observational value derived in this study.
The number of bright \h2 regions decreases as ages increase. 
Fitting the bright part ($L >  10^{37.0}$\ergs) yields 
$\alpha=-2.35$, --2.39, and --2.35 for age = 0, 2, and 4 Myr, respectively.
There are few bright \h2 regions in the case of age = 6 Myr,
but a fit including the faint part ($ L > 10^{36.0}$\ergs) yields a similar index, $\alpha=-2.35$. 
Thus the values of $\alpha$ change little depending on age, and they are almost the same as the input value of $\beta$.

\subsection{A Change in Slope of \ha Luminosity Function for \h2 Regions and the Initial Cluster Mass Function}

A controversial point in the \h2 LF for M51 has been a break at $L \approx 10^{38.8}$\ergs~shown in previous studies
based on the ground-based imaging \citep{ken89,ran92,thi00}. 
\citet{ken89} argued that the cause of this break might be a transition from the normal to the super giant \h2 regions.
On the other hand, the break is possibly a feature caused by the low spatial resolution.
\citet{sco01} compared their \h2 LF of M51 derived from $HST$ WFPC2 with that of \citet{ran92} based on the ground-based images.
They demonstrated that the high-resolution \h2 LF is significantly steeper than the low-resolution one,
and showed that the break at $L \approx 10^{38.8}$\ergs~is not seen in their \h2 LF.

\begin{figure}
    \plotone{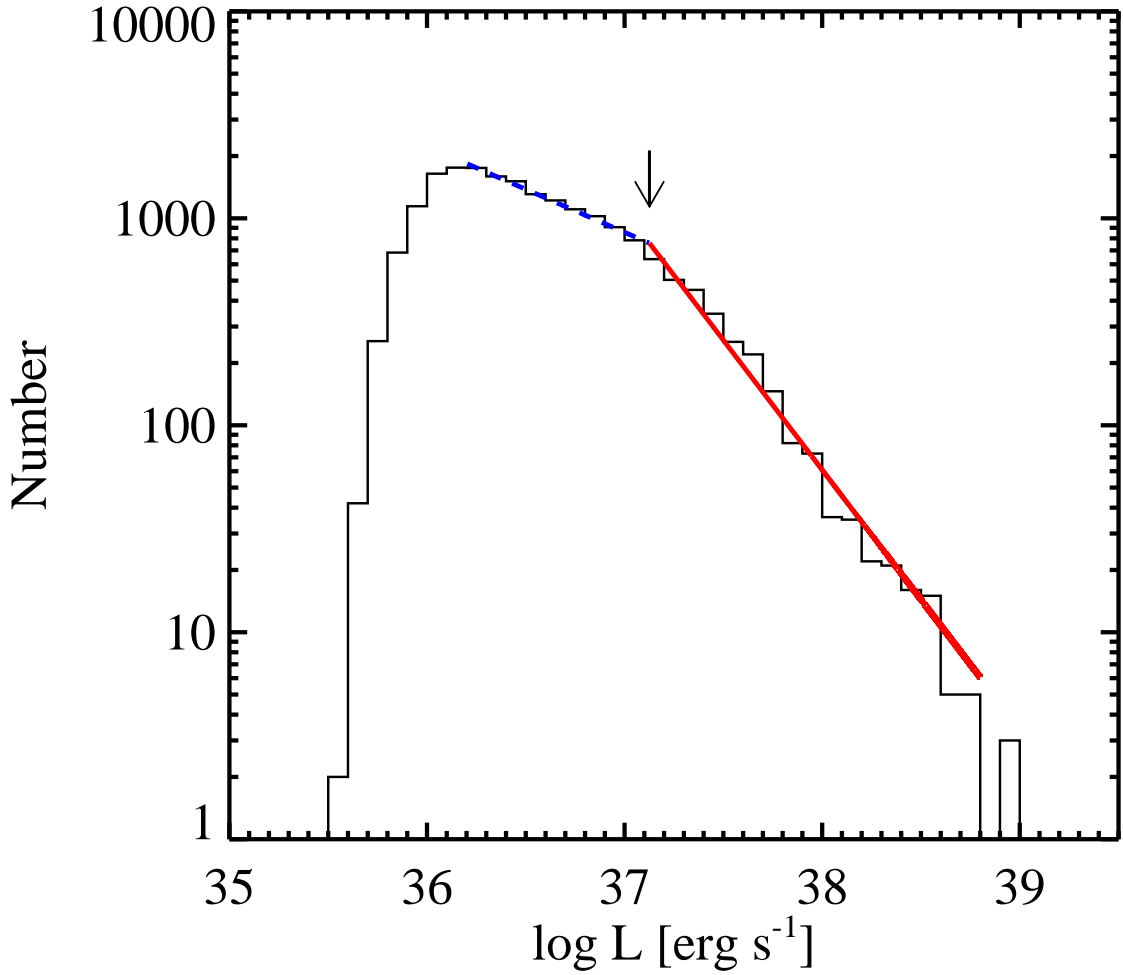}
\caption{The \h2 LF for the original sample (histogram).
Thick dashed and thick solid lines represent a double power law fit for the faint part ($L < 10^{37.1}$\ergs) 
and the bright part ($L > 10^{37.1}$\ergs), respectively.
The power law indices are $\alpha=-2.25\pm0.02$ for the bright part and  $\alpha=-1.42\pm0.01$ for the faint part.
An arrow marks the break point.}
\label{fig3}
\end{figure}

\begin{figure}
    \epsscale{0.6}
    \plotone{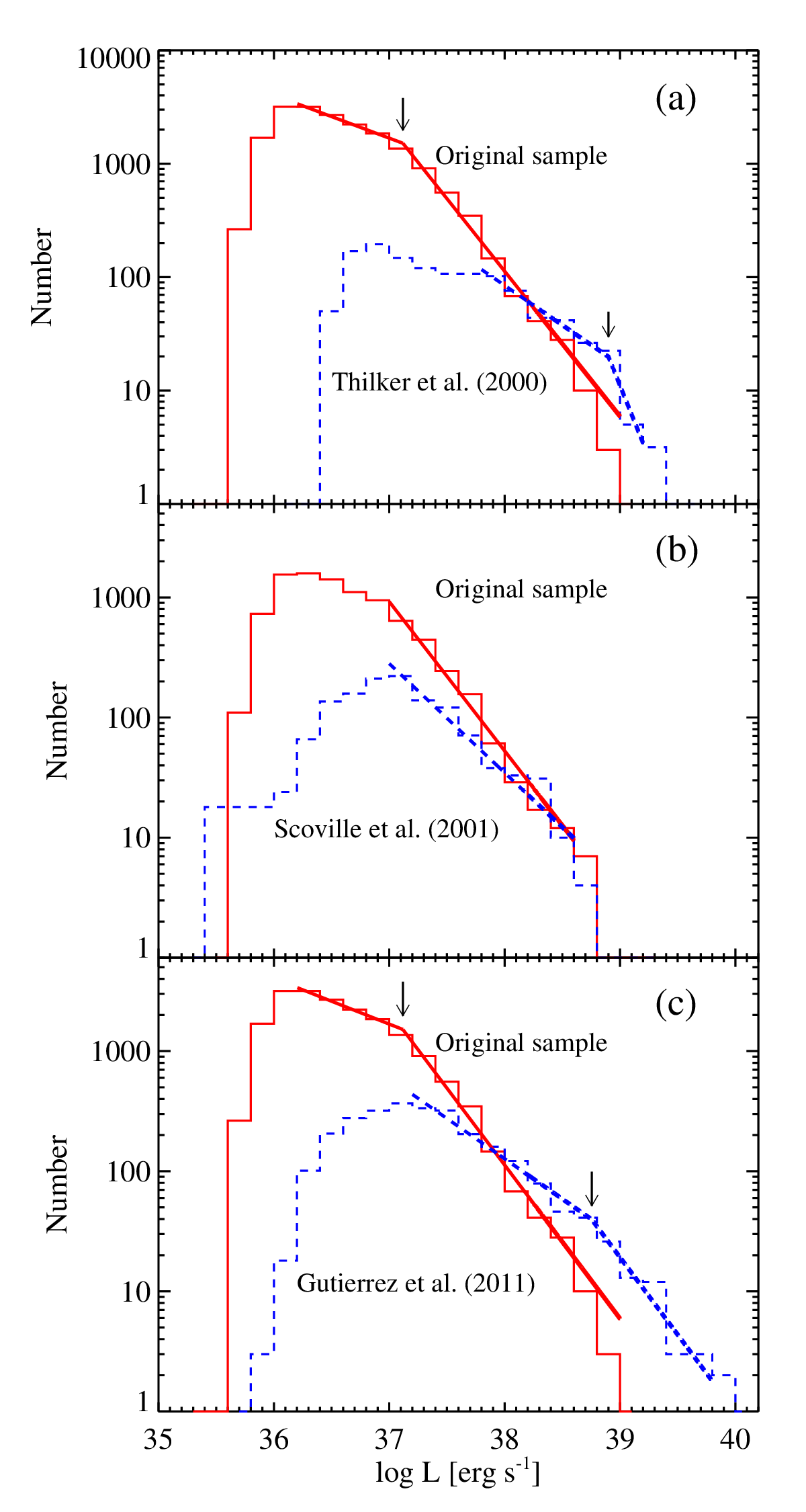}
\caption{Comparison of the \h2 LFs  for the original sample (solid line)
with those in \citet{thi00} (dashed line in (a)), in \citet{sco01} (dashed line in (b)) and  in \citet{gut11} (dashed line in (c)).
Thick lines represent power law fits.
In panel (a) and (c), this study shows a break at $L= 10^{37.1}$\ergs, 
while \citet{thi00} and \citet{gut11} show a break at $L=10^{38.8}$\ergs, much brighter than that in this study.
Arrows mark the break point.}
\label{fig4}
\end{figure}

\begin{figure}
    \epsscale{0.6}
    \plotone{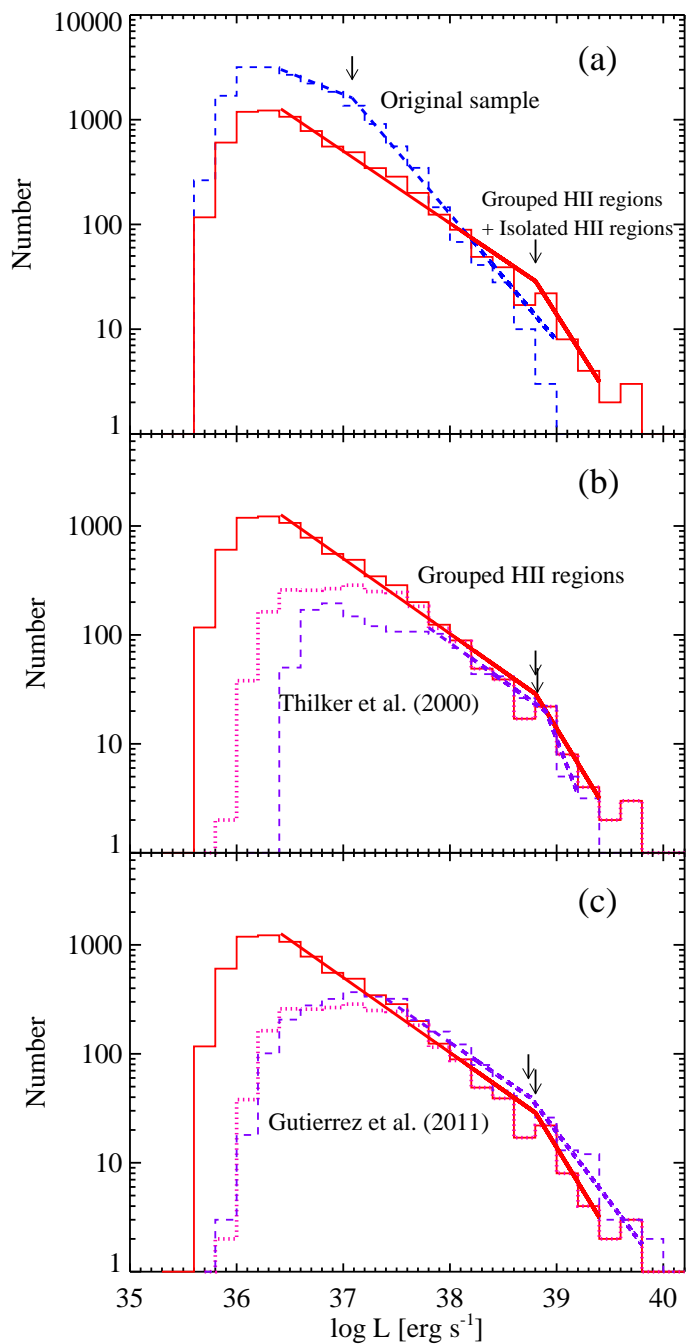}
\caption{(a) The \h2 LFs for the original (dashed line) and group (solid line) samples.
(b) The \h2 LFs for the group sample (solid line), the grouped \h2 regions (dotted line) 
and in \citet{thi00} (dashed line). 
(c) The \h2 LFs for the group sample (solid line), the grouped \h2 regions (dotted line) 
and in \citet{gut11} (dashed line). 
Thick lines represent power law fits. Arrows mark the break point.}
\label{fig5}
\end{figure}

\begin{figure}
    \epsscale{0.6}
    \plotone{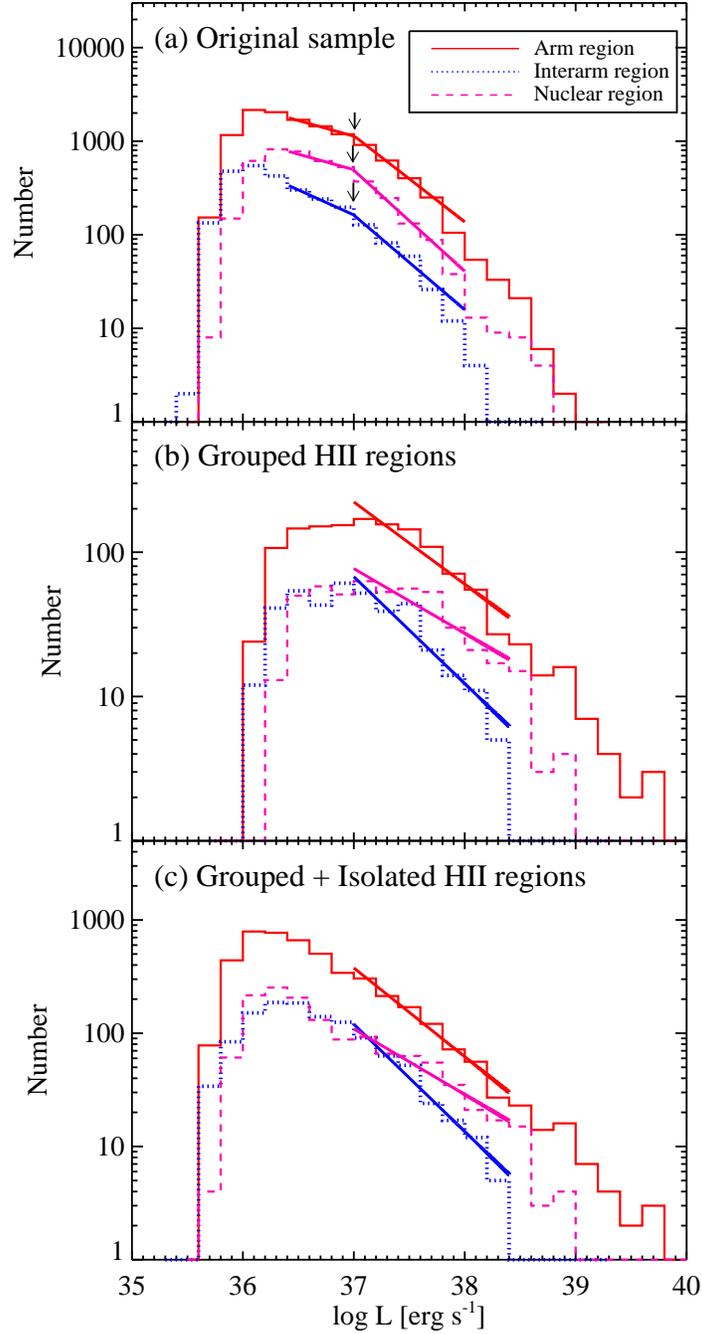}
\caption{(a) The \h2 LFs for the arm (solid line), interarm (dotted line), and nuclear (dashed line) regions in the original sample.
(b) The \h2 LFs for the arm (solid line), interarm (dotted line), and nuclear (dashed line) regions in the grouped \h2 regions.
(c) The \h2 LFs for the arm (solid line), interarm (dotted line), and nuclear (dashed line) regions in the group sample.
Thick solid lines represent power law fits. 
Arrows mark the break point.}
\label{fig6}
\end{figure}

\begin{figure}
    \plotone{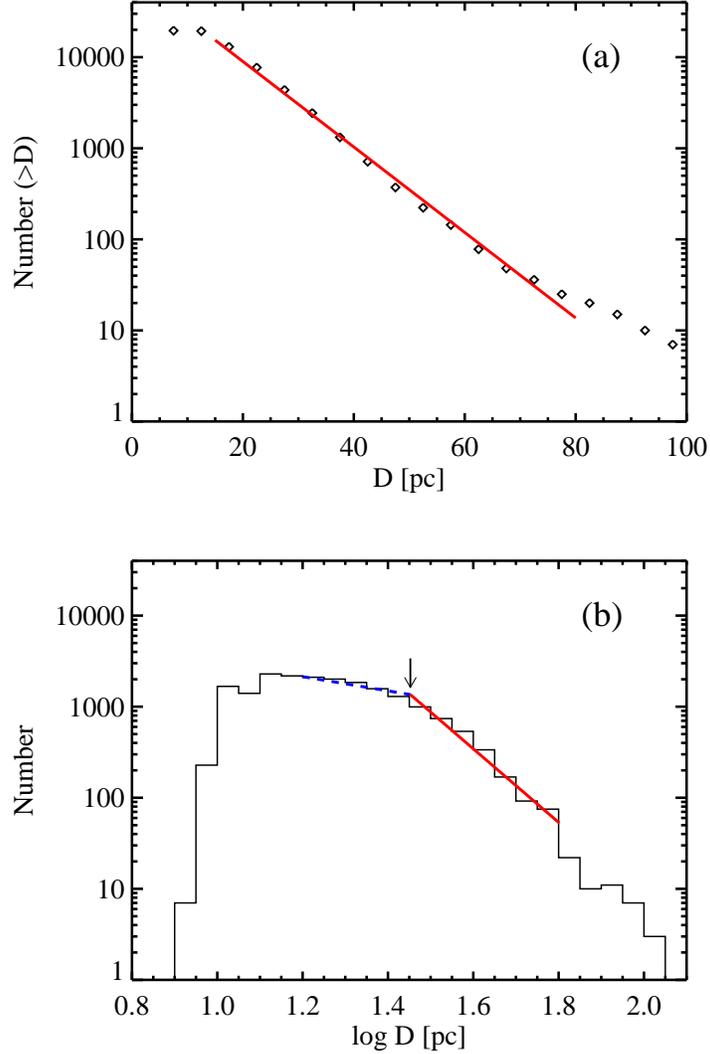}
\caption{The size distribution of the \h2 regions in the original sample.
(a) Cumulative size distribution. The solid line represents an exponential fit.
(b) Differential size distribution. 
Thick dashed and thick solid lines represent a double power law fit with a break at log $D = 1.45$ ($D\approx30$ pc).
The power law index for small \h2 regions with $D < 30$ pc is $\alpha_D = -5.04\pm0.08$,
whereas $\alpha_D = -1.78\pm0.04$ for large \h2 regions with $D > 30$ pc.
An arrow marks the break point.}
\label{fig7}
\end{figure}

\begin{figure}
    \plotone{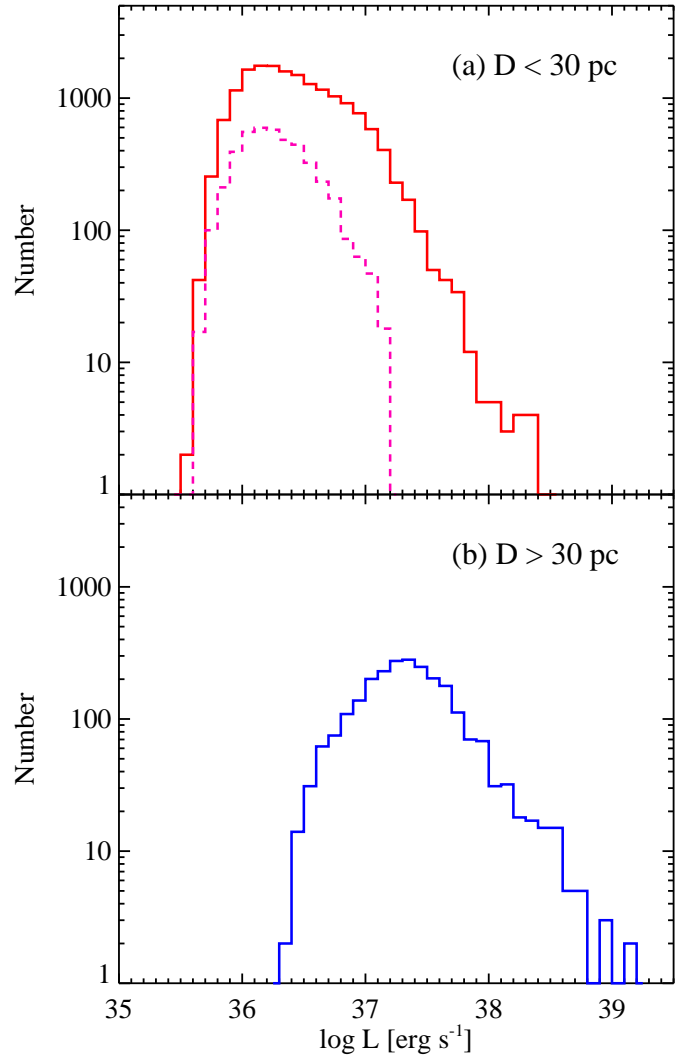}
\caption{(a) The \h2 LFs for the small \h2 regions with $D < 30$ pc (solid line)
and the isolated \h2 regions which are neither blended with neighbor sources nor located in the crowded regions (dashed line).
(b) The \h2 LF for the large \h2 regions with $D > 30$ pc.}
\label{fig8}
\end{figure}

\begin{figure}
    \plotone{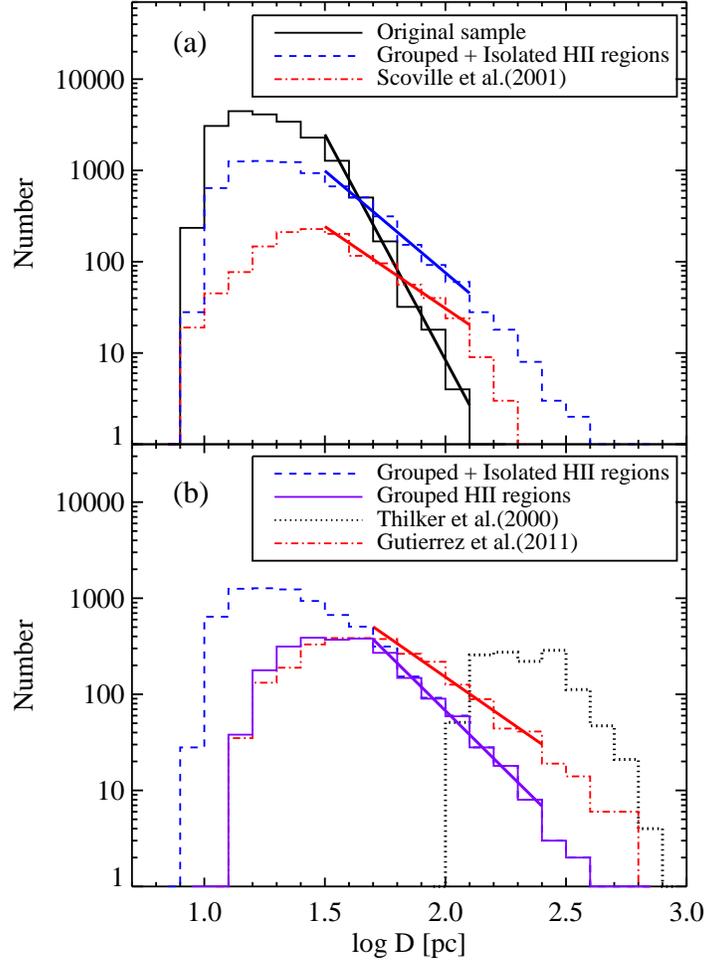}
\caption{(a) The size distributions of the \h2 regions in the original (solid line) and group (dashed line) samples 
in comparison with those in \citet{sco01} (dot-dashed line). Thick solid lines represent power law fits for 1.5 $<$ log $D <$ 2.1.
(b) The size distributions for the grouped \h2 regions (solid line) and the \h2 regions in the group sample (dashed line)  
in comparison with those in \citet{thi00} (dotted line) and in \citet{gut11} (dot-dashed line).
Thick solid lines represent power law fits for 1.7 $<$ log $D <$ 2.4.}
\label{fig9}
\end{figure}

\begin{figure}
    \epsscale{0.85}
    \plotone{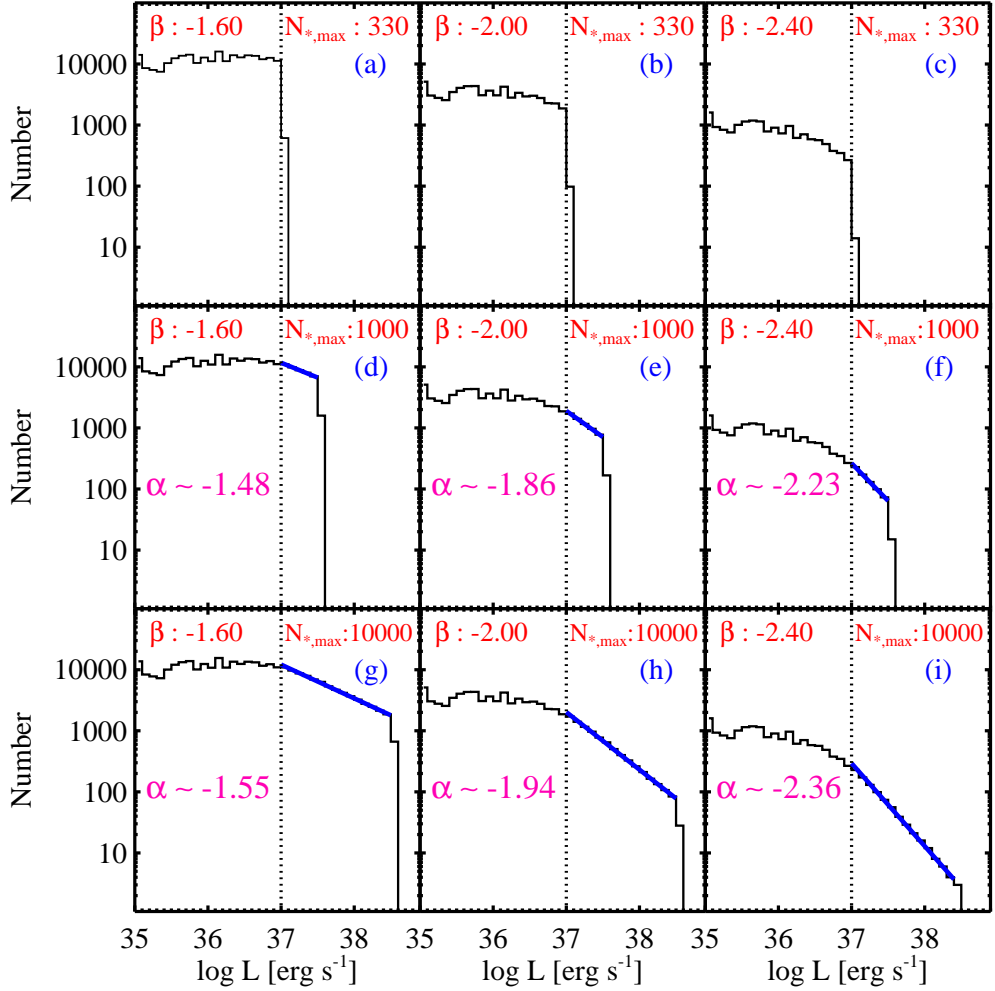}
\caption{Simulated \ha LFs of the \h2 regions with zero age for three parent power law indices ($\beta$ = --1.60, --2.00, and --2.40) 
and for the three maximum number of stars per cluster ($\rm N_{*,max}$ = 330, 1,000, and 10,000).
Thick solid lines in (d) to (i) represent power law fits for the bright part ($L > 10^{37.0}$\ergs).
The values of the index ($\alpha$) are given in each panel.}
\label{fig10}
\end{figure}

\begin{figure}
    \plotone{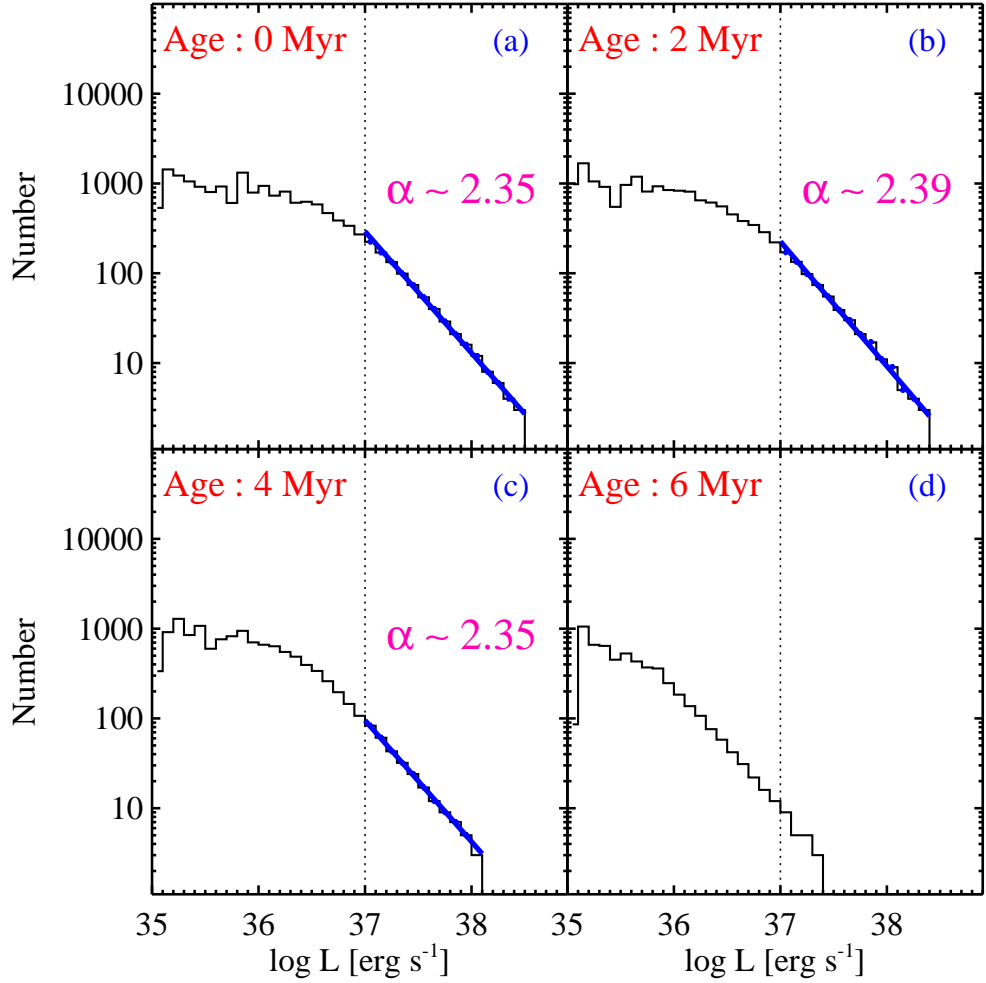}
\caption{Evolution of simulated \h2 LFs for four ages (0, 2, 4, and 6 Myr),
the parent power law index $\beta$ = --2.40, and the maximum number of stars per cluster $\rm N_{*,max}$ = 10,000.
Thick solid lines represent power law fits for $L > 10^{37.0}$\ergs.
The values of the index ($\alpha$) are given in each panel.}
\label{fig11}
\end{figure}

This break is not seen either in the \h2 LF for the original sample in this study, as shown in Figure \ref{fig3}.
However, we can observe a break at $L = 10^{38.8}$\ergs~in the \h2 LF for the group sample in Figure \ref{fig5}.
From the similarity between this \h2 LF and that in \citet{thi00} in Figure \ref{fig5}(b),
we conclude that the break seen at $L \approx 10^{38.8}$\ergs~in the previous studies
may be due to the blending of multiple \h2 regions.

On the other hand, we found a break at $L = 10^{37.1}$\ergs~in the \h2 LF 
for the  original sample as shown in Figure \ref{fig3}.
To understand the shape of the \h2 LF with this break 
we compared the observational \h2 LF for the original sample with the simulated \h2 LF in Figure \ref{fig10}.
The shape of the observational \h2 LF is remarkably similar to that of the simulated \h2 LF  in Figure \ref{fig10}(i) 
derived for $\beta = -2.40$ and $\rm N_{*,max}$ = 10,000.
The observational \h2 LF shows a break at $L = 10^{37.1}$\ergs,
which is consistent with the simulated \h2 LF with a break at $L=10^{37.0}$\ergs.   
The power law indices of the observational \h2 LF are  $\alpha=-2.25\pm0.02$ 
for the bright part and $\alpha=-1.42\pm0.01$ for the faint part,
agreeing with those in the simulated \h2 LF, $\alpha=-2.36\pm0.03$ and $\alpha=-1.43\pm0.02$.
Thus it is concluded that the break at $L=10^{37.1}$\ergs~is
due to the different $\rm N_*$ in clusters associated with individual \h2 regions.
In other words, the faint \h2 regions with $L < 10^{37.1}$\ergs~are ionized by low mass clusters with
$\rm N_{*} < 330$ (including several OB stars) or single massive stars,
while the bright \h2 regions with $L > 10^{37.1}$\ergs~
are ionized by massive clusters with  $\rm N_{*} >> 330$ (including several tens of OB stars).

It is noted that the cluster mass $\rm M_{cl}$ is related with $\rm N_*$: $\rm M_{cl} \propto \rm N_*$.
Therefore  the power law index $\alpha$ of \h2 LF in the bright end
is expected to be similar to the index  $\beta$ (--2.25) of  the IMF of the clusters in M51 that  
may be represented by a power law: $\rm N(M_{cl})~dM_{cl}={M_{cl}}^{-2.25}dM_{cl}$.
This slope ($\beta$) is similar to or slightly steeper than those derived from previous studies for M51 young clusters:
from the observation of young star clusters in M51, 
the cluster mass function is known to have power law form with its index 
ranging from $\alpha=-1.70\pm0.08$ \citep{gie06} to $\alpha=-2.23\pm0.34$  \citep{hwa10}.

\subsection{Spatial Variation of \ha Luminosity Function}

Figure \ref{fig6} shows that the shape of the \h2 LFs varies depending on the location within a galaxy.
In Figure \ref{fig6}(a) for the original sample, we found that 
although the \h2 LFs for the bright part ($L > 10^{37}$\ergs) have similar power law indices in three regions,
the \h2 LF for the faint part ($L < 10^{37}$\ergs) is steeper in the interarm region than in the arm and nuclear regions.
In addition, there is no \h2 region brighter than $L = 10^{38.2}$\ergs~in the interarm region.
In Figure \ref{fig6}(b) and (c) for the grouped \h2 regions and the group sample,
we also found that the \h2 LF is steeper in the interarm region than in the arm and nuclear regions.

Although the shape of the \h2 LF can be affected by several factors, including variation in the metallicity,
extinction, stellar IMF, and underlying gas dynamics, 
\citet{ken89} argued that the most significant factor is the change in the total number of stars produced in
typical star formation events. 
\citet{ran92} attributed the difference between the arm and the interarm \h2 LF 
to the existence of a more massive molecular clouds in the arm region.
In other words, the higher proportion of massive clouds gives rise to relatively more \h2 regions of the higher luminosity.
It is reasonable to infer that
the high gas surface densities combined with the compression by the strong density wave in the arms
may cause a concentration of bright \h2 regions on the arm region.
Later \citet{oey98} suggested that the interarm \h2 regions are on average fainter than those in the arm region 
because the ionizing sources in the arm region are younger than those in the interarm region.

In Figure \ref{fig11}, we show the evolution of \h2 LFs with $\beta =$ --2.40 and $\rm N_{*,max}$ = 10,000
for different ages (0, 2, 4, and 6 Myr).
The \h2 LFs shift to faintward as a function of time.
A maximum luminosity after 4 Myr (panel c) is 0.5 dex fainter than at zero-age (panel a), 
and the power law index for the faint part ($L < 10^{37.1}$\ergs) after 4 Myr is steeper than at zero-age.
The interarm \h2 LF in Figure \ref{fig6} has a lower maximum luminosity 
and a steeper slope for the faint part than the arm \h2 LF.
Thus the difference between the arm and the interarm \h2 LFs  can be explained by an effect of evolution,
supporting the suggestion by \citet{oey98}. 

\subsection{Relation between Luminosities and Sizes of \h2 Regions}

Both the \ha luminosity function and  the size distribution of \h2 regions in the original sample 
are fitted by a double power law as shown in Figure \ref{fig3} and \ref{fig7}(b), respectively.
Since the total \ha luminosity of an \h2 region is integrated from the nebular volume emission, 
it is expected that  $L \propto {D}^{3}$. 
Then the power law indices (${\alpha}$ and ${\alpha}_{D}$) of the \ha luminosity function and  the size distribution for \h2 regions
are related as ${\alpha}_{D}= 2 - 3{\alpha}$ \citep{oey03}.
 
The \h2 LF for the original sample in Figure \ref{fig3} shows that the break appears at $L = 10^{37.1}$\ergs,
and that the power law indices are $\alpha=-1.42\pm0.01$ for the faint part and $\alpha=-2.25\pm0.02$ for the bright part.
Using these values we derived expected values, $\alpha_D =-2.26$ and $\alpha_D=-4.75$ for the faint and bright part, respectively. 
On the other hand, the observational size distribution in Figure \ref{fig7}(b) shows that 
the power law index for the small \h2 regions  with $D < 30$ pc  is  $\alpha_D =-1.78\pm0.04$,
whereas $\alpha_D =-5.04\pm0.10$ for the large \h2 regions with $D > 30$ pc. 
Therefore there is a good agreement between expected and observational indices 
for the size distribution of \h2 regions.

\begin{figure}
    \epsscale{0.6}
    \plotone{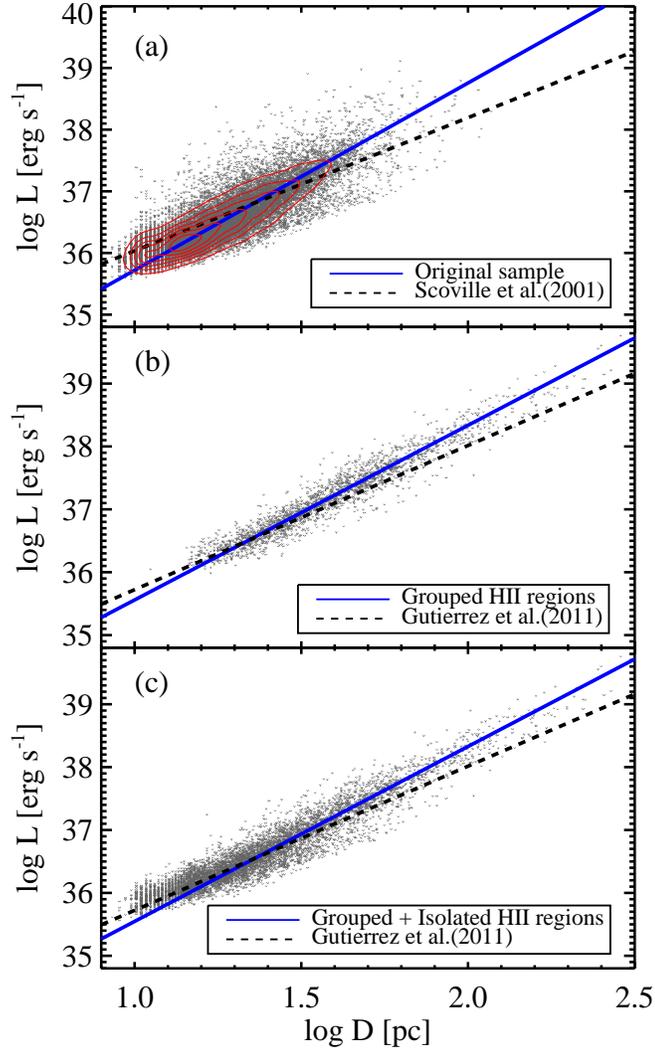}
\caption{(a) The \ha luminosities versus sizes of the \h2 regions in the original sample (dots and contours) and in \citet{sco01}(dashed line).
Contours represent the number density of \h2 regions in the original sample.
The solid line represents a power law fit: $L\propto{D}^{3.04\pm0.01}$,
and the dashed line represents a fit to the data in \citet{sco01}: $L\propto{D}^{2.16\pm0.02}$.
(b) The \ha luminosities versus sizes of the grouped \h2 regions (dots) and the \h2 regions in \citet{gut11}(dashed line).
The solid line represents a power law fit: $L\propto{D}^{2.78\pm0.02}$,
and the dashed line represents a fit to the data in \citet{gut11}: $L\propto{D}^{2.29\pm0.03}$.
(c) The \ha luminosities versus sizes of the \h2 regions in the group sample (dots) and in \citet{gut11}(dashed line).
The solid line represents a power law fit: $L\propto{D}^{2.78\pm0.01}$,
and the dashed line represents a fit to the data in \citet{gut11}: $L\propto{D}^{2.29\pm0.03}$.}
\label{fig12}
\end{figure}

Figure \ref{fig12}(a) displays \ha luminosities versus sizes of the \h2 regions in the original sample.
It shows a clear correlation between the two parameters.
The relation between the luminosity and the size of \h2 regions is fitted by $L \propto {D}^{3.04\pm0.01}$ (solid line).
This value is in good agreement with our expectation of $L \propto {D}^{3}$.
It is noted, however, \citet{sco01} derived a flatter relation,
$L\propto{D}^{2.16\pm0.02}$ for \h2 regions in the central region of M51 (dashed line).
They explained that the this flatter slope is due to blending effect:
large \h2 regions are blended so that they include substantial empty or neutral volumes.
This indicates that this study separated blended \h2 regions better than \citet{sco01}.

%%%
Figure \ref{fig12}(b) and (c) display \ha luminosities versus sizes of the grouped \h2 regions and
the \h2 regions in the group sample, respectively.
The relation between the luminosity and the size of \h2 regions for the two samples
is fitted by $L \propto {D}^{2.78\pm0.02}$ (panel b; solid line) and $L \propto {D}^{2.78\pm0.01}$(panel c; solid line).
\citet{gut11} derived a flatter relation, $L\propto{D}^{2.29\pm0.03}$  (dashed line).
This discrepancy between the two studies is due to the different procedure 
used to derive the size of the \h2 region, as mentioned in \S5.3.

\section{Summary and Conclusion}

We detected and analysed the \h2 regions in M51 using the wide field and high resolution $HST$ ACS \ha image 
taken  part of the Hubble Heritage program. 
We found about 19,600 \h2 regions using the automatic \h2 photometry software, HIIPHOT \citep{thi00}.
Primary results are summarized as follows.

The \ha luminosity of the \h2 regions ranges from $L$ = $10^{35.5}$\ergs~to $10^{39.0}$\ergs.
The \h2 LF is fitted by a double power law with a break at $L = 10^{37.1}$\ergs,
and the power law indices are $\alpha=-2.25\pm0.02$ for the bright part and $\alpha=-1.42\pm0.01$ for the faint part.
Comparison with the simulated \h2 LFs suggests that this break 
is caused by the transition of \h2 region ionizing sources, 
from low mass clusters (including several OB stars) to massive clusters (including several tens of OB stars).

To understand the break in the \h2 LF at $L = 10^{38.8}$\ergs~reported in the ground-based studies,
we produced a group sample, finding associations of \h2 regions in the original sample using the FOF\citep{dav85}.
From the comparison of the \h2 LF for the group sample with that of \citet{thi00},
we concluded that most \h2 regions brighter than $L = 10^{38.8}$\ergs~seen in the ground-based studies
may be blends of multiple lower luminosity \h2 regions that are resolved as separate \h2 regions in this study.

The \h2 LFs for the original sample have a different shape according to their positions in the galaxy.
Although the \h2 LFs are fitted by a double power law with a break at $L$ = $10^{37.1}$\ergs, 
the \h2 LF for the faint part ($L < 10^{37.1}$\ergs) is steeper in the interarm region than in the arm and nuclear regions.
There is no \h2 region brighter  than $L = 10^{38.2}$\ergs~in the interarm region.
Observed variations in the \h2 LF can be explained by an effect of evolution of an \h2 region.

The size distribution of the \h2 regions in the original sample 
shows that the size of \h2 regions ranges from 8 pc to 110 pc in diameter.
The cumulative and differential size distributions of the \h2 regions 
are fitted well by an exponential and power law form, respectively.
The cumulative size distribution of the \h2 regions with 15 pc $< D <$ 80 pc is fitted well 
with an exponential law: $N(D)=N_{0}$ exp $(-D/D_{0})$ with $N_{0}=77,880$ and $D_{0}=9.3$ pc.

The differential size distribution of these \h2 regions 
is fitted by a double power law with a break at $D=30$ pc.
The power law index for the small \h2 regions with 15 pc $< D <$ 30 pc is $\alpha_D = -1.78\pm0.04$,
whereas $\alpha_D = -5.04\pm0.08$ for the large \h2 region with 30 pc $< D <$ 110 pc.
The small \h2 regions with $D < 30$ pc may be  ionized  by several OB stars. 
On the other hand, the large \h2 regions with $D > 30$ pc are considered to be superpositions of small \h2 regions,
and they can be associated with stellar associations involving several tens of OB stars.
In addition, power law indices of the size distribution for the original sample are related with those of \h2 LF, 
and the relation between the luminosities and sizes of \h2 regions is fitted well by $L\propto{D}^{3.04\pm0.01}$.

%\acknowledgement

We thank David Thilker for providing his \h2 region photometry software HIIPHOT 
and the \h2 region catalog for M51.
We are also grateful to Nick Scoville and Leonel Gutierrez for sharing a catalog for the \h2 regions in M51.
M.G.L. was supported by Mid-career Researcher Program through NRF grant funded by the MEST (No.2010-0013875).
This research was supported by the BK21 program of the Korean Government.
N.H. acknowledges the support by Grant-in-Aid for JSPS Fellow No. 20-08325.

\clearpage

\clearpage

\clearpage

\clearpage

\clearpage

\clearpage

\clearpage

\clearpage

\clearpage

\clearpage

\begin{deluxetable}{ccccccccc}
\tabletypesize{\normalsize}
\setlength{\tabcolsep}{0.005in}
\tablecaption{A catalog of M51 \h2 regions in the original sample \label{table1}}
\tablewidth{0pt}
\tablehead{
\colhead{\footnotesize{~ID}} & \colhead{\footnotesize{~~R.A.(J2000.0)}\tablenotemark{a}~} & \colhead{\footnotesize{~Decl.(J2000.0)}\tablenotemark{a}~} 
& \colhead{\footnotesize{F(H$\alpha$)}\tablenotemark{b} } & \colhead{\footnotesize{~log L(H$\alpha$)}\tablenotemark{c}}  
& \colhead{~D\tablenotemark{d}~}   & \colhead{~~D~~}  & \colhead{~\footnotesize{Pos.\tablenotemark{e}~}} &  \colhead{\footnotesize{GID}\tablenotemark{f}} \\
& \colhead{\footnotesize{[Deg]}}       & \colhead{\footnotesize{[Deg]}}      &    & \colhead{\footnotesize{[erg $s^{-1}$]}}   & \colhead{\footnotesize{[\arcsec]}}  &
\colhead{\footnotesize{~[pc]}} &   & }
\startdata
  3 &  202.3956299 &     47.171803   ~&    4.402~~& 36.573 &  0.53   ~&  25.26  &      I ~&        \\   
    4 &  202.3961639 &     47.169731   ~&    7.095~~& 36.780 &  0.44   ~&  21.32  &      I ~&        \\   
    6 &  202.3963776 &     47.169849   ~&    1.259~~& 36.029 &  0.34   ~&  16.47  &      I ~&        \\   
    7 &  202.3977661 &     47.172359   ~&    8.074~~& 36.836 &  0.57   ~&  27.22  &      I ~&        \\   
    8 &  202.3983917 &     47.172512   ~&    1.597~~& 36.133 &  0.38   ~&  18.37  &      A ~&        \\   
    9 &  202.3992462 &     47.177734   ~&    1.669~~& 36.152 &  0.38   ~&  18.37  &      A ~&     10      \\
   10 &  202.3993530 &     47.177658   ~&    1.290~~& 36.040 &  0.28   ~&  13.27  &      A ~&     10      \\
   11 &  202.3993835 &     47.177444   ~&   13.960~~& 37.074 &  0.79   ~&  38.11  &      A ~&     10      \\
   13 &  202.4014740 &     47.182098   ~&   13.420~~& 37.057 &  0.58   ~&  27.75  &      A ~&        \\   
   20 &  202.4026184 &     47.160851   ~&    4.296~~& 36.562 &  0.68   ~&  32.83  &      A ~&        \\   
\enddata
\tablenotetext{a}{\small{Measured in the F658N (H$\alpha$) image.}}
\tablenotetext{b}{\small{In the unit of [$10^{-16}$ erg~$\rm cm^{-2}$~$\rm s^{-1}$].}}
\tablenotetext{c}{\small{Calculated using $L=4\pi \rm {d}^{2}\times $F(H$\alpha$)  for d=9.9 Mpc.}}
\tablenotetext{d}{\small{Calculated using $D=2\times( \rm area/\pi)^{1/2}$ where the area is derived from the number of pixels contained in an \h2 region.}}
\tablenotetext{e}{\small{Position in the galaxy: A--Arm region, 
I--Interarm region, N--Nuclear region, and C--NGC 5195.}}
\tablenotetext{f}{\small{ID number for the group sample.}}
\end{deluxetable}

\begin{deluxetable}{cccrcccc}
\tabletypesize{\normalsize}
\setlength{\tabcolsep}{0.005in}
\tablecaption{A catalog of M51 \h2 regions in the group sample\label{table2}}
\tablewidth{0pt}
\tablehead{
\colhead{\footnotesize{~ID}} & \colhead{\footnotesize{~~R.A.(J2000.0)}\tablenotemark{a}~} & \colhead{\footnotesize{~Decl.(J2000.0)}\tablenotemark{a}~} 
& \colhead{\footnotesize{F(H$\alpha$)}\tablenotemark{b} } & \colhead{\footnotesize{~log L(H$\alpha$)}\tablenotemark{c}}  
& \colhead{~D\tablenotemark{d}~}   & \colhead{~~D~~}  & \colhead{~\footnotesize{Pos.\tablenotemark{e}~}} \\
& \colhead{\footnotesize{[Deg]}}       & \colhead{\footnotesize{[Deg]}}      &    & \colhead{\footnotesize{[erg $s^{-1}$]}}   & \colhead{\footnotesize{[\arcsec]}}  &
\colhead{\footnotesize{~[pc]}} &  }
\startdata
    3 &  202.3956299 &     47.171803   ~&    4.402~~& 36.573 &  0.53   ~&  25.26  &    I~\\
    4 &  202.3961639 &     47.169731   ~&    7.095~~& 36.780 &  0.44   ~&  21.32  &    I~\\
    6 &  202.3963776 &     47.169849   ~&    1.259~~& 36.029 &  0.34   ~&  16.47  &    I~\\
    7 &  202.3977661 &     47.172359   ~&    8.074~~& 36.836 &  0.57   ~&  27.22  &    I~\\
    8 &  202.3983917 &     47.172512   ~&    1.597~~& 36.133 &  0.38   ~&  18.37  &    A~\\
   10 &  202.3993530 &     47.177654   ~&   17.928~~& 37.183 &  0.96   ~&  45.96  &    A~\\
   13 &  202.4014740 &     47.182098   ~&   13.420~~& 37.057 &  0.58   ~&  27.75  &    A~\\
   16 &  202.4025269 &     47.178154   ~&    1.830~~& 36.192 &  0.34   ~&  16.25  &    A~\\
   17 &  202.4025269 &     47.178005   ~&    1.471~~& 36.097 &  0.34   ~&  16.25  &    A~\\
   20 &  202.4026184 &     47.160851   ~&    4.296~~& 36.562 &  0.68   ~&  32.83  &    A~\\

\enddata
\tablenotetext{a}{\small{Measured in the F658N(H$\alpha$) image. Mean R.A.(J2000) and Decl.(J2000) of all individual \h2 regions in a group.}}
\tablenotetext{b}{\small{In the unit of [$10^{-16}$ erg~$\rm cm^{-2}$~$\rm s^{-1}$]. Derived from summing the \ha fluxes of all individual \h2 regions in a group.}}
\tablenotetext{c}{\small{Calculated using $L=4\pi\rm d^2\times $F(H$\alpha$) for d=9.9 Mpc.}}
\tablenotetext{d}{\small{Calculated using $D=2\times( \rm area/\pi)^{1/2}$
where the area is derived from summing the numbers of pixels contained in all individual \h2 regions in a group.}}
\tablenotetext{e}{\small{Position in the galaxy: A--Arm region, 
I--Interarm region, N--Nuclear region, and C--NGC 5195.}}
\end{deluxetable}
\clearpage

\end{document}